\documentclass[sn-basic]{sn-jnl}

\usepackage{graphicx}%
\usepackage{multirow}%
\usepackage{amsmath,amssymb,amsfonts}%
\usepackage{amsthm}%
\usepackage{mathrsfs}%
\usepackage[title]{appendix}%
\usepackage{xcolor}%
\usepackage{textcomp}%
\usepackage{manyfoot}%
\usepackage{booktabs}%
\usepackage{algorithm}%
\usepackage{algorithmicx}%
\usepackage{algpseudocode}%
\usepackage{listings}%
\usepackage{bm}
\usepackage{lineno}

\theoremstyle{thmstyleone}

\theoremstyle{thmstyletwo}%

\theoremstyle{thmstylethree}%

\begin{document}

\title[Article Title]{Discovery of physically interpretable wave equations}

\author*[1]{\fnm{Shijun} \sur{Cheng}}\email{sjcheng.academic@gmail.com}

\author[1]{\fnm{Tariq} \sur{Alkhalifah}}\email{tariq.alkhalifah@kaust.edu.sa}

\affil*[1]{\orgdiv{Division of Physical Science and Engineering}, \orgname{King Abdullah University of Science and Technology}, \orgaddress{\city{Thuwal}, \postcode{23955-6900}, \country{Saudi Arabia}}}

\abstract{Using symbolic regression to discover physical laws from observed data is an emerging field. In previous work, we combined genetic algorithm (GA) and machine learning to present a data-driven method for discovering a wave equation. Although it managed to utilize the data to discover the two-dimensional $(x,z)$ acoustic constant-density wave equation $u_{tt}=v^2(u_{xx}+u_{zz})$ (subscripts of the wavefield, $u$, are second derivatives in time and space) in a homogeneous medium, it did not provide the complete equation form, where the velocity term is represented by a coefficient rather than directly given by $v^2$. In this work, we redesign the framework, encoding both velocity information and candidate functional terms simultaneously. Thus, we use GA to simultaneously evolve the candidate functional and coefficient terms in the library. Also, we consider here the physics rationality and interpretability in the randomly generated  potential wave equations, by ensuring that both-hand sides of the equation maintain balance in their physical units. We demonstrate this redesigned framework using the acoustic wave equation as an example, showing its ability to produce physically reasonable expressions of wave equations from noisy and sparsely observed data in both homogeneous and inhomogeneous media. Also, we demonstrate that our method can effectively discover wave equations from a more realistic observation scenario.}


\maketitle

\section*{Article Highlights}
\begin{itemize}
   \item We encode both functional and coefficient terms, enabling the simultaneous evolution of their potential forms. \\
   \item  We leverage the critical characteristic that physical equations must satisfy unit balance to restrict the generation of non-physical equations. \\
  \item  Our method can effectively discover a physically interpretable wave equation from sparse, noisy, and highly realistic observed wavefields in inhomogeneous media. \\
  
\end{itemize}

\section{Introduction}
The wave equation is classified as a partial differential equation (PDE). PDEs are essential to our day-to-day simulation of natural phenomena, like wave propagation \citep{kjartansson1979constant, thomsen1986weak, carcione1990wave, alkhalifah2000acoustic}. Establishing an accurate physical equation is crucial for studying a dynamic system. It enables us to predict the behavior of the system under different conditions, thus understanding how the system responds to various internal and external factors. For example, Newton's formulation of universal gravitation allows astronomers to predict the motion of planets and other celestial bodies. Typically, mathematical equations are derived from established physical laws, which in turn, are grounded in empirical evidence collected from rigorous scientific experiments. 

In realm of equations describing wave propagation, which are dictated by well-known hyperbolic PDEs, they are derived from Newton's second law and Hooke's linear theory \citep{shearer2019introduction}. While the resulting PDEs often explain our observations, there are cases where the collected data do not conform to established laws, or phenomena that have not yet been physically described. For example, 
wave equations corresponding to velocity anisotropy are widely used in seismic modeling \citep{thomsen1986weak, tsvankin1997anisotropic, xu2020new}, inversion \citep{warner2013anisotropic, alkhalifah2014recipe, oh2020multistage}, and imaging \citep{mu2020least, liang2020scattering, liang2022born, ouyang2023multiparameter}. However, experimental measurements suggest that attenuation anisotropy may be more pronounced than velocity anisotropy \citep{hosten1987inhomogeneous, zhu2006plane, zhu2007plane, cheng2021plane, wang2022propagating}. The Biot equation \citep{biot1956theory1, biot1956theory2, biot1962generalized} has been proposed to study the seismic wave propagation in a solid medium saturated with fluid. However, its estimates for attenuation and dispersion are significantly lower than experimental measurements \citep{dvorkin1993dynamic, ba2017rock, liu2020seismic, cheng2021wave}.
These equations rely on certain assumptions about the Earth's properties. However, the Earth, as it vibrates, often acts beyond the bounds imposed by these assumptions. Under these circumstances, it seems logical to seek a new, more accurate mathematical equation grounded in existing knowledge and physical principles to replace the original wave equation. However, this process can be intricate and time-intensive.

Benefiting from the recent advances in machine learning (ML) and data-processing capabilities, data-driven discovery methods have been developed to identify the underlying PDEs of physical problems \citep{champion2019data, lejarza2022data, tenachi2023deep}. Contrary to the paradigm where physical laws are deduced based on set physical principles, the data-driven approach for discovering equations offers an alternative perspective. This approach identifies the governing equations of dynamical systems directly from observed data, which might be noisy \citep{xu2019dl, reinbold2021robust}. Such a method could be practical for systems with intricate underlying mechanisms. Essentially, data-driven discovery methods entail creating a library of candidate functional terms, followed by employing diverse optimization algorithms to determine the optimal combination, thereby generating the general form of equations. Therefore, we can see that the construction of a library plays a pivotal role in data-driven discovery of PDEs.

Currently, library construction methods are typically categorized as closed or expandable \citep{chen2022symbolic}. Closed methods involve initially building an overcomplete library, then utilizing sparse regression techniques, such as Lasso \citep{schaeffer2017learning}, sequential threshold ridge regression \citep{rudy2017data}, and SINDy \citep{brunton2016discovering}, to determine dominant candidate functional terms. However, these methods are limited to predetermined complete candidate libraries, making it challenging to ensure the inclusion of true PDE terms, particularly without prior knowledge. As candidate libraries expand, sparsifying them becomes increasingly difficult, raising the risk of identifying incorrect PDEs. Meanwhile, derivative computation from observations often relies on methods like finite difference or polynomial interpolation, which are vulnerable to irregular measurement grids and noise. In contrast, expandable library methods offer greater potential for identifying PDEs with complex structures by starting with a randomly generated incomplete initial library, which evolves through genetic algorithms (GA) to produce unlimited combinations \citep{maslyaev2019data}. Despite such advancements, handling noisy and sparse data remains a challenge. Some researchers have turned to neural network (NN)-based functional representation to compute derivatives using automatic differentiation, offering stability and robustness to noisy data compared to traditional numerical methods \citep{xu2023discovery}.

In previous work \citep{cheng2023d, cheng2024robust}, we tested such discovery algorithms, with our own flavor of implementation, in directly discovering the wave equation from observed spatial-temporal wavefields. Specifically, we first pre-train an NN to approximate the relationship between the spatial-temporal coordinates ($x$, $y$, $z$, $t$) and the observed wavefield, denoted as $u(x, y, z, t)$. This pre-trained NN is then utilized to interpolate observed wavefield, and also, calculate the time and spatial derivatives of the wavefield, such as $u_{tt}$ and $u_{xx}$, through automatic differentiation. Subsequently, a selection process involving a GA refines an initially extensive set of potential terms for the wave equation into a more focused preliminary library. This step is followed by employing a physics-informed criterion (PIC) \citep{xu2023discovery} to assess the accuracy and parsimony of the potential equations, thereby discovering the optimal wave equation structure. In the final phase, a physics-informed NN (PINN) \citep{raissi2019physics}, now embedded with the discovered wave euqation, is trained to identifing the precise coefficients for each functional term in the equation.

With the dual support of NNs and GA, we demonstrated that our algorithm can discover the wave equation from pressure wavefields in a homogeneous medium. However, we emphasize that we did not provide a complete form. That is, the velocity term in the wave equation is represented by a coefficient optimized by the physics-informed neural network (PINN). For example, when given a homogeneous medium with velocity of 2 $\texttt{km}/\texttt{s}$, the equation form discovered from synthesized pressure wavefields (given by $u$) is $u_{tt} = 3.999(u_{xx} + u_{zz})$, rather than $u_{tt} = v^2(u_{xx} + u_{zz})$. Here, subscripts of the wavefield $u$ denotes derivatives in space $(x,z)$ and time ($t$), and specifically second derivatives. We can see that although 3.999 is close to the value of $v^2=4 \texttt{km}^2/\texttt{s}^2$, the resulting equation lacks balance in the physical units on both-hand sides (BHS). For instance, the unit on the left-hand side (LHS) is $\texttt{km}/\texttt{s}^2$ (considering the wavefield is given by displacement, as an example), while on the right-hand side (RHS), it is $1/\texttt{km}$. This leads to a discovered equation lacking physical interpretability. Moreover, another important issue is that the discovered equation is only valid for homogeneous media in which the observed wavefield are collected. Although it seems that we have discovered the wave equation and the corresponding velocity from the observed data, this was valid for homogeneous media only. When there are variations in the medium velocity, the value of the predicted single velocity value becomes uninterpretable (maybe be some average).

To address these issues, we, here, modify our algorithm to enable it to discover physically interpretable wave equations. Specifically, we assume that velocity is measurable and encode it along with the functional terms. In so doing, each functional term is paired with a coefficient term, which is given by an encoded velocity. This enables us to utilize GA to evolve both the candidate functional terms and their corresponding coefficient terms, simultaneously. Then, when generating candidate functional and velocity terms, we impose a physical unit constraint to ensure unit consistency on the BHS of the potential equation. This effectively prevents the generation of non-physical candidate terms. We test the effectiveness of our method in discovering the 2-D acoustic wave equation, where seismic wave propagates in both homogeneous and inhomogeneous media, and also, demonstrate its robustness to noisy and sparse observed data. Furthermore, we consider a more realistic observation system, where the wavefields are only collected from the top surface of the model, to demonstrate the discovery capabilities of our method under very limited observations.

The rest of paper is structured as follows. We first review the original data-driven framework for discovering wave equations. Then, we detail the improvements made to the original algorithm, including how to encode coefficient terms, impose physical unit constraints, and determine the wave equations. Subsequently, we share numerical examples to illustrate the process of discovering a physically interpretable equation using the new framework. Meanwhile, we demonstrate its robustness to noise and sparse observations, and show its capability to discover wave equations from wavefields collected in heterogeneous media and realistic observation systems. Furthermore, we discuss the role of physical unit constraints in data-driven discovery of wave equation and outline future research directions. Finally, we summarize this work and draw some conclusions.

\section{Review of data-driven discovery of wave equation}
\begin{figure}[!t]
\centering
\includegraphics[width=1\textwidth]{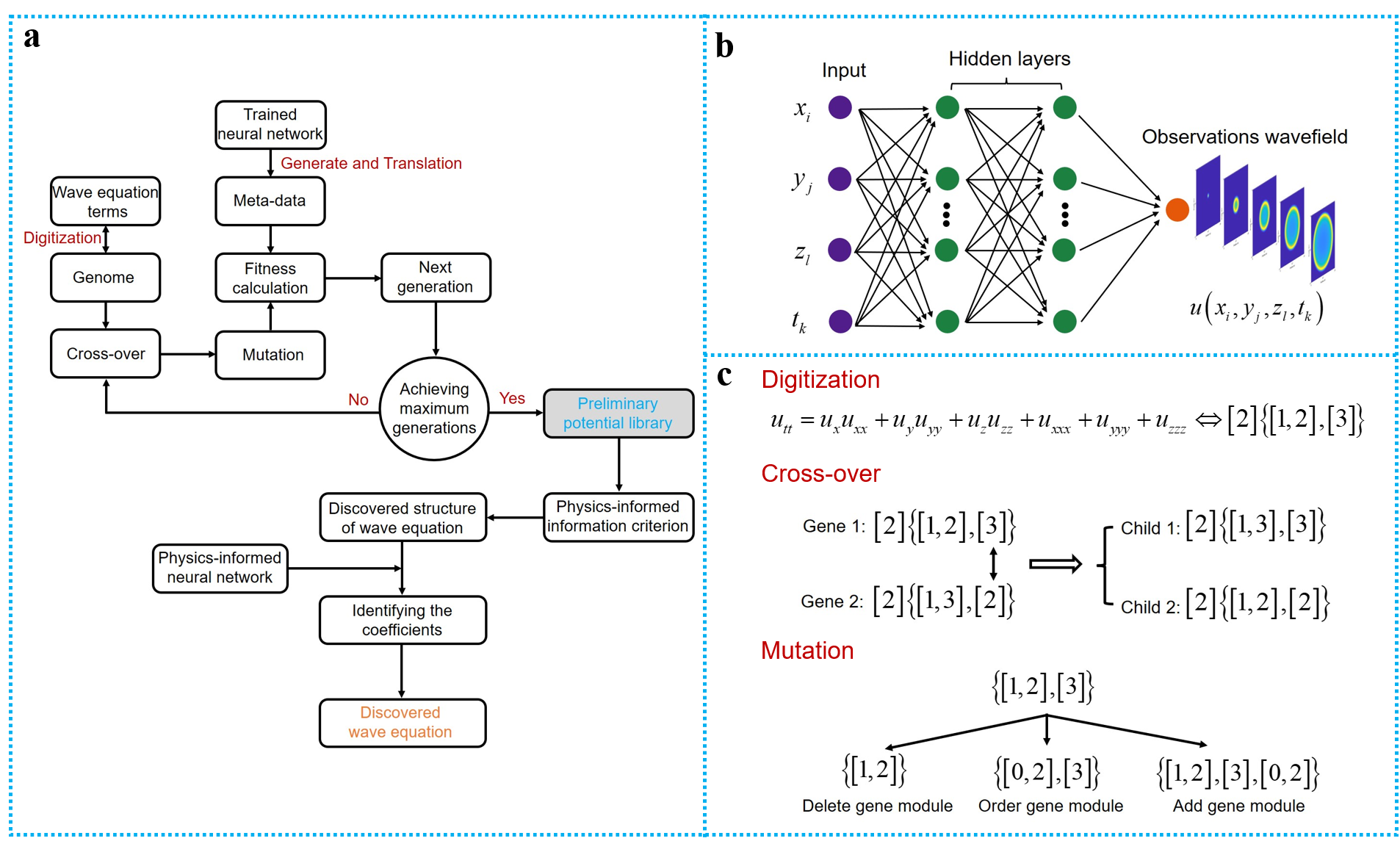}
\caption{An illustration of the data-driven discovery of a seismic wave equation. (a) Implementation workflow with machine learning and genetic algorithm. (b) Structure of the deep fully connected back-propagation neural network used to approximate the observation wavefield. (c) An illustration of digitization, cross-over, and mutation of genetic algorithm.}
\label{fig1}
\end{figure}

Our original framework, dubbed D-WE, consisted of two components: the neural network (NN) and the genetic algorithm (GA). We illustrate the original workflow in Figure \ref{fig1}a. For a potential wave equation, the LHS of the equation typically includes only the time derivative terms, while the RHS contains the spatial derivative terms and their corresponding coefficients. Therefore, we consider the general form of the wave equation as
\begin{equation}\label{eq1}
u_{T} = f\left (\boldsymbol{\Theta}(u);\left[\xi_{i}\right]_{i=1,\cdot\cdot\cdot,n} \right)
\end{equation}
with
\begin{equation}\label{eq2}
\boldsymbol{\Theta}(u)=\left[u, u_x, u_y, u_z, u_{x x}, u_{y y}, u_{z z}, \cdots\right],
\end{equation}
where ${\textit u}_T$ represents different orders of time derivatives of the displacement or pressure wavefield ${\textit u}$ (in this work we will assume $u$ represents displacement with units of space, like meters) with respect to time ${\textit t}$, e.g., first ($u_t$) or second ($u_{tt}$); ${\Theta}(u)$ refers to the library composed of candidate functional terms, in which the subscripts denote different orders of spatial derivatives; $\left[\xi_i \right]_{i=1,\cdot\cdot\cdot,n}$ is the vector of coefficients with size $n$ of the candidates functional terms in the library; and $f\left(\cdot \right)$ is a function parameterizing a wave equation with possible contributing functional terms. 

When observing the seismic wavefield, discretized as $u\left (x_i,y_j,z_k,t_l \right)$, $i=1,\cdot\cdot\cdot, N_x$, $j=1,\cdot\cdot\cdot, N_y$, $k=1,\cdot\cdot\cdot, N_z$, and $l=1,\cdot\cdot\cdot, N_t$, we can concretize the discovery problem in equation \ref{eq1} as follows:
\begin{equation}\label{eq3}
\left[\begin{array}{c}
u_T(\textbf{x}, t)_1 \\
u_T(\textbf{x}, t)_2 \\
\ldots \\
u_T(\textbf{x}, t)_N
\end{array}\right]=\left[\begin{array}{ccc}
u(\textbf{x}, t)_1 & u_x(\textbf{x}, t)_1 & \cdots \\
u(\textbf{x}, t)_2 & u_x(\textbf{x}, t)_2 & \ldots \\
\ldots & \ldots & \ldots \\
u(\textbf{x}, t)_N & u_x(\textbf{x}, t)_N & \ldots
\end{array}\right]\left[\begin{array}{c}
\xi_1 \\
\ldots \\
\xi_n
\end{array}\right].
\end{equation}
where $\textbf{x}=(x_i,y_j,z_k)$ denote the spatial locations of observations, the subscript index of $(\textbf{x}, t)$ represents the ordinal number of the spatial-temporal observed point, and $N$ represents the number of total observations and equals to $N_x \cdot N_y \cdot N_z \cdot N_t$. Therefore, the objective of data-driven discovery of the wave equation is to solve a large linear system of equations represented by equation \ref{eq3}. Typically, we can employ optimization algorithms to obtain the non-zero coefficients from the entire coefficient set $\left[\xi_i \right]_{i=1,\cdot\cdot\cdot,n}$, while the functional terms corresponding to zero coefficients are removed \citep{brunton2016discovering, schaeffer2017learning, rudy2017data}. As a result, we can select real functional terms from the library ${\Theta}(u)$, and also, determine the coefficients corresponding to each functional term.

To achieve this goal, we first train an NN to approximate the nonlinear relationship between the spatial-temporal coordinates and the observed wavefields (see Figure \ref{fig1}b), which serves to compute spatial and time derivatives and generate metadata. Then, we present a specific encoding technique to represent the potential wave equation as its corresponding genome (see Figure \ref{fig1}c). Specifically, different order of time or spatial derivatives in a potential wave equation $u_{tt} = u_xu_{xx} + u_yu_{yy} + u_zu_{zz} + u_{xxx} + u_{yyy} + u_{zzz}$ can be represented by the following encoding:
\begin{equation}\label{eq4}
\text{Gene:} \left\{\begin{array}{l}
{0 \Leftrightarrow u} \\
{1 \Leftrightarrow u_x \quad \text{or} \quad u_y \quad \text{or} \quad u_z \quad \text{or} \quad u_t} \\
{2 \Leftrightarrow u_{xx} \quad \text{or} \quad u_{yy} \quad \text{or} \quad u_{zz} \quad \text{or} \quad u_{tt}} \\
{3 \Leftrightarrow u_{xxx} \quad \text{or} \quad u_{yyy} \quad \text{or} \quad u_{zzz}} \\
\end{array}\right..
\end{equation}
Here, we refer to the numbers 0, 1, 2, and 3 as genes. By combining genes, we can form a gene module to represent a functional term. For example, the first and second-order derivatives on the LHS of the equation can be represented by gene modules [1] and [2], respectively. For the RHS of the equation, the two functional terms $u_xu_{xx} + u_yu_{yy} + u_zu_{zz}$ and $u_{xxx} + u_{yyy} + u_{zzz}$ shown in Figure \ref{fig1}c can be represented by $[1, 2]$ and $[3]$, respectively. The representation of the entire RHS of the equation can be combined into $\{[1, 2], [3]\}$. By combining the digitized representations of LHS of the equation, we can obtain a genome corresponding to a potential equation, such as $[2]\{[1, 2], [3]\}$.

After encoding the potential equations, we randomly generate a population containing a predefined number of genomes. In our case, this population contains 400 genomes, which correspond to 400 potential equations. This population undergoes crossover and mutation to produce the next generation. Then, we measure the fitness of each genome. Based on the calculated fitness scores, we select the best half of the genomes as the next generation, while the other half is replaced by new random genomes. We repeat the process of crossover, mutation, and selection to evolve the population. When reaching the maximum predefined number of iterations, the best genome in the population becomes a preliminary library. However, this preliminary library may still contain some redundant functional terms. Thus, we further refine it using more accurate physics-informed information criterion (PIC) \citep{xu2023discovery} to determine the optimal structure of the wave equation. Once we obtain the general structure of the wave equation, we need to determine the coefficients corresponding to each functional term on the RHS of the equation. In the original implementation, we relied on the discovered structure of the wave equation, treating the coefficients as learnable parameters. We utilized PINN to optimize the parameters, resulting in more accurate coefficient values. All the descriptions above constitute the main workflow of our original framework. More details can be found in \cite{cheng2024robust}. 

However, we can see that the original framework cannot determine the specific form of coefficient terms. Instead, it only used PINN to obtain the coefficient values for each functional term without considering physical units. This imbalance in physical units between the LHS and RHS of the discovered equation lead to a lack of physical interpretability. As a result, the discovered equation work solely in the homogeneous medium where the wavefield are observed, rendering it inapplicable to alternative media. Hence, in the subsequent section, we will enhance our algorithm to discover a physically interpretable wave equation.
\section{Method}
\subsection{Problem statement}
To simultaneously discover the complete form of both functional and the coefficient terms, and also, provide a physically interpretable equation, we need to reconsider the general form of a wave equation as
\begin{equation}\label{eq5}
u_{T} = f\left (\boldsymbol{\Theta}(u);\boldsymbol{\Theta}(v) \right)
\end{equation}
with
\begin{equation}\label{eq6}
\boldsymbol{\Theta}(u)=\left[u, u_x, u_y, u_z, u_{x x}, u_{y y}, u_{z z}, \cdots\right], 
\boldsymbol{\Theta}(v)=\left[1, v, v^2, v^3, \cdots\right],
\end{equation}
where ${\Theta}(v)$ represents a new library composed of candidate coefficient terms, assuming here the only parameter representing the medium is velocity. So, we only consider encoding velocity and assume that the velocity is known. In practical scenarios, we often rely on experimental measurements of medium velocity, like from wells. This can also be utilized in cases where we build physical inhomogeneous models in a lab in which we know the medium properties, like velocities, and we use the physical model to discover a wave equation from observed data. 

Specifically, for an observed wavefield, equation \ref{eq3} can be re-expressed as
\begin{equation}\label{eq7}
\left[\begin{array}{c}
u_T(\textbf{x}, t)_1 \\
u_T(\textbf{x}, t)_2 \\
\ldots \\
u_T(\textbf{x}, t)_N
\end{array}\right]=\left[\begin{array}{ccc}
v(\textbf{x})_1 u(\textbf{x}, t)_1 & v^2(\textbf{x})_1 u_x(\textbf{x}, t)_1 & \cdots \\
v(\textbf{x})_2 u(\textbf{x}, t)_2 & v^2(\textbf{x})_2 u_x(\textbf{x}, t)_2 & \ldots \\
\ldots & \ldots & \ldots \\
v(\textbf{x})_N u(\textbf{x}, t)_N & v^2(\textbf{x})_N u_x(\textbf{x}, t)_N & \ldots
\end{array}\right]\left[\begin{array}{c}
\xi_1 \\
\ldots \\
\xi_n
\end{array}\right].
\end{equation}
We can see that the original objective (e.g., equation \ref{eq3}) only required determining the values of coefficients, where the non-zero coefficients in the coefficient set $\left[\xi_i \right]_{i=1,\cdot\cdot\cdot,n}$ served as the coefficients preceding each functional term. Obviously, in complex wave physics equations, an equation with constant coefficients can only serve as a special case for a homogeneous medium. In contrast, here, each functional term corresponds to a coefficient term, which has an explicit form instead of a constant value. The coefficient set $\left[\xi_i \right]_{i=1,\cdot\cdot\cdot,n}$ to be determined thus becomes a dimensionless parameter set. Consequently, by employing optimization algorithms, we can determine the non-zero coefficients from the dimensionless parameter set, along with the corresponding functional terms and paired coefficient terms also becomes established.

Definitely, this poses a significant challenge. It implies that we need to optimize two function libraries ${\Theta}(u)$ and ${\Theta}(v)$ simultaneously. In the original framework, we developed an encoding scheme for potential functional terms and then utilized a GA algorithm to optimize library ${\Theta}(u)$, thus obtaining a preliminary library. For the upgraded objective here, in the next section, we will illustrate how to couple the optimization of library ${\Theta}(v)$ into the original framework.

\subsection{Digitization of coefficient terms}
To utilize GA for evolving library ${\Theta}(v)$, we design a similar digitization criterion for encoding coefficient terms. Specifically, the different power of $v$ are represented as follows:
\begin{equation}\label{eq8}
{0 \Leftrightarrow 1 }, ~~ {1 \Leftrightarrow v^2 }, ~~  {2 \Leftrightarrow v^2 }, ~~ {3 \Leftrightarrow v^3 }, ~~ \text{and} ~~ {4 \Leftrightarrow v^4 },
\end{equation}
where the number 0, 1, 2, 3, and 4 denote the genes. Here, the highest power of $v$ is considered as 4. Higher powers are not common to most wave equations. 

In the original digitization framework, we use combinations of genes to form a gene module, which represents a functional term. For example, the functional term $(uu_{xxx} + uu_{yyy} + uu_{zzz})$ can be represented as gene module $[0,3]$, where gene 0 represents $u$, and gene 3 denotes the spatial derivatives $u_{xxx}$, $u_{yyy}$, and $u_{zzz}$, respectively. Similarly, we can use gene modules to represent a coefficient term, for instance, 
\begin{equation}\label{eq9}
{[0] \Leftrightarrow 1 }, ~~ {[1] \Leftrightarrow v^2 }, ~~  {[2] \Leftrightarrow v^2 }, ~~ {[3] \Leftrightarrow v^3 }, ~~ \text{and} ~~ {[4] \Leftrightarrow v^4 }.
\end{equation}
We can see that equations \ref{eq8} and \ref{eq9} are quite similar, with the only difference being the addition of square brackets enclosing a coefficient term in equation \ref{eq9}. For each functional term on the RHS of the equation, we must match a corresponding coefficient term encoding. That is, each functional term is paired with a corresponding coefficient term. For example, for the form $v^2(uu_{xxx} + uu_{yyy} + uu_{zzz})$, its functional term encoding is $[0,3]$, and the paired coefficient term encoding is $[2]$. 

For a potential equation, we represent it as a genome, which consists of many gene modules. For example,  
\begin{equation}\label{eq10}
u_{tt} = v^2(u_{xx} + u_{yy} + u_{zz}) + v^2(uu_{xxx} + uu_{yyy} + uu_{zzz}) \Leftrightarrow \text{Genome:} [2]\{[2], [0, 3]\}\{[2], [2]\},
\end{equation}
where the first gene module $[2]$ corresponds to the encoding of the LHS of the equation, and the combination of the gene module $\{[2], [0, 3]\}$ represents the encoding of the functional terms on the RHS of equation, with their corresponding coefficient term encoding denoted as $\{[2], [2]\}$. Here, on the LHS of the equation, we consider only the first- and second-order time derivatives, as we did in the original framework. This already covers the majority of wave equations.

\subsection{Physical unit constraints}
In the previous section, we established corresponding encoding scheme for the coefficient terms, allowing us to generate potential equations that simultaneously include both the functional terms and their corresponding coefficient terms. However, within the context of physics, there's an important principle that must be met: the generated equations must be physically interpretable. A key component of physically interpretable equations is that the physical units on BHS of the equation is balanced \citep{tenachi2023deep}. Any equation that merely fits numerical values without satisfying unit conservation is generally meaningless.

In the population generation and evolution process, to ensure that each potential equation is physically interpretable, we introduce unit constraints to prevent the generation of non-physical potential equations. Firstly, we propose a rule for physical unit encoding. As we know, the typical wave equation involves units such as meters ($\texttt{m}$) and seconds ($\texttt{s}$), neglecting density. Here, we count the units of $\texttt{m}$ and $\texttt{s}$ separately. For the LHS of the equation, we only consider first and second-order time derivatives, while for the functional terms on the RHS, we consider up to third-order spatial derivatives, and for the coefficient terms, we only consider up to $v^4$. For the unit $\texttt{m}$, we assign the following numbers to each partial derivatives and coefficient term encoding:
\begin{equation}\label{eq11}
\left\{\begin{array}{l}
{-2 \Leftrightarrow u_{xxx} \quad \text{or} \quad u_{yyy} \quad \text{or} \quad u_{zzz}} \\
{-1 \Leftrightarrow u_{xx} \quad \text{or} \quad u_{yy} \quad \text{or} \quad u_{zz}} \\
{0 \Leftrightarrow u_x \quad \text{or} \quad u_y \quad \text{or} \quad u_z} \\
{1 \Leftrightarrow u \quad \text{or}\quad u_t  \quad \text{or}\quad u_{tt} \quad \text{or} \quad v} \\
{2 \Leftrightarrow v^2 } \quad \quad  {3 \Leftrightarrow v^3 } \quad \quad  {4 \Leftrightarrow v^4 }\\
\end{array}\right..
\end{equation}
The reason we define it this way is based on the units of each quantity involved. For example, for $u_{xxx}$, its physical unit is $1/\texttt{m}^2$, thus we represent it as $-2$. For the unit $\texttt{s}$, the case is simpler, since there are only the time derivatives and velocity terms involved. Hence, their corresponding encodings are defined as:
\begin{equation}\label{eq12}
{1 \Leftrightarrow u_{t} ~~\text{or}~~v }, ~~ {2 \Leftrightarrow u_{tt} ~~ \text{or} ~~ v^2 }, ~~  {3 \Leftrightarrow v^3 }, ~~ \text{and} ~~ {4 \Leftrightarrow v^4 }.
\end{equation}

Once we establish the units for each quantity, we define a rule for calculating the physical units. For the LHS of the equation, the calculation is straightforward. If the LHS involves a first-order time derivative, both $\texttt{m}$ and $\texttt{s}$ have units of 1. If it involves a second-order time derivative, the units for $\texttt{m}$ and $\texttt{s}$ are 1 and 2, respectively. For the RHS of the equation, the calculation includes adding the corresponding numbers for the units of each quantity involved in the multiplication of spatial derivatives and each functional term with its corresponding coefficient term. For example, for $v^2(uu_{xxx} + uu_{yyy} + uu_{zzz})$, the units for $\texttt{m}$ and $\texttt{s}$ corresponding to coefficient term $v^2$ are both 2, while for functional term $(uu_{xxx} + uu_{yyy} + uu_{zzz})$, the unit for $\texttt{m}$ is -1. As a result, the units for $\texttt{m}$ and $\texttt{s}$ in form $v^2(uu_{xxx} + uu_{yyy} + uu_{zzz})$ are 1 and 2, respectively. 

According to this calculation rule, we can evaluate whether the generated genomes, which represent potential equations, are balanced in terms of physical units. For example, for equation $u_{tt} = v^2(u_{xx} + u_{yy} + u_{zz}) + v^2(uu_{xxx} + uu_{yyy} + uu_{zzz})$, the units for $\texttt{m}$ and $\texttt{s}$ on the LHS are 1 and 2, respectively, and on the RHS, the units for $\texttt{m}$ and $\texttt{s}$ are also 1 and 2. Therefore, the physical units on BHS of this equation are balanced, indicating a physical model. For equation $u_{t} = v^2(u_{xx} + u_{yy} + u_{zz}) + v^2(uu_{xxx} + uu_{yyy} + uu_{zzz})$, the unit for $\texttt{s}$ on the LHS is 1, which does not match the unit for $\texttt{s}$ on the RHS, making it a non-physical model. When randomly generated potential equations have unbalanced units, we do not include them in the libraries ${\Theta}(u)$ and ${\Theta}(v)$. Instead, we discard them. Therefore, in evolving the population using GA, all genomes are physical models. 

\subsection{Determination of wave equation}
\begin{figure}
\centering
\includegraphics[width=1\textwidth]{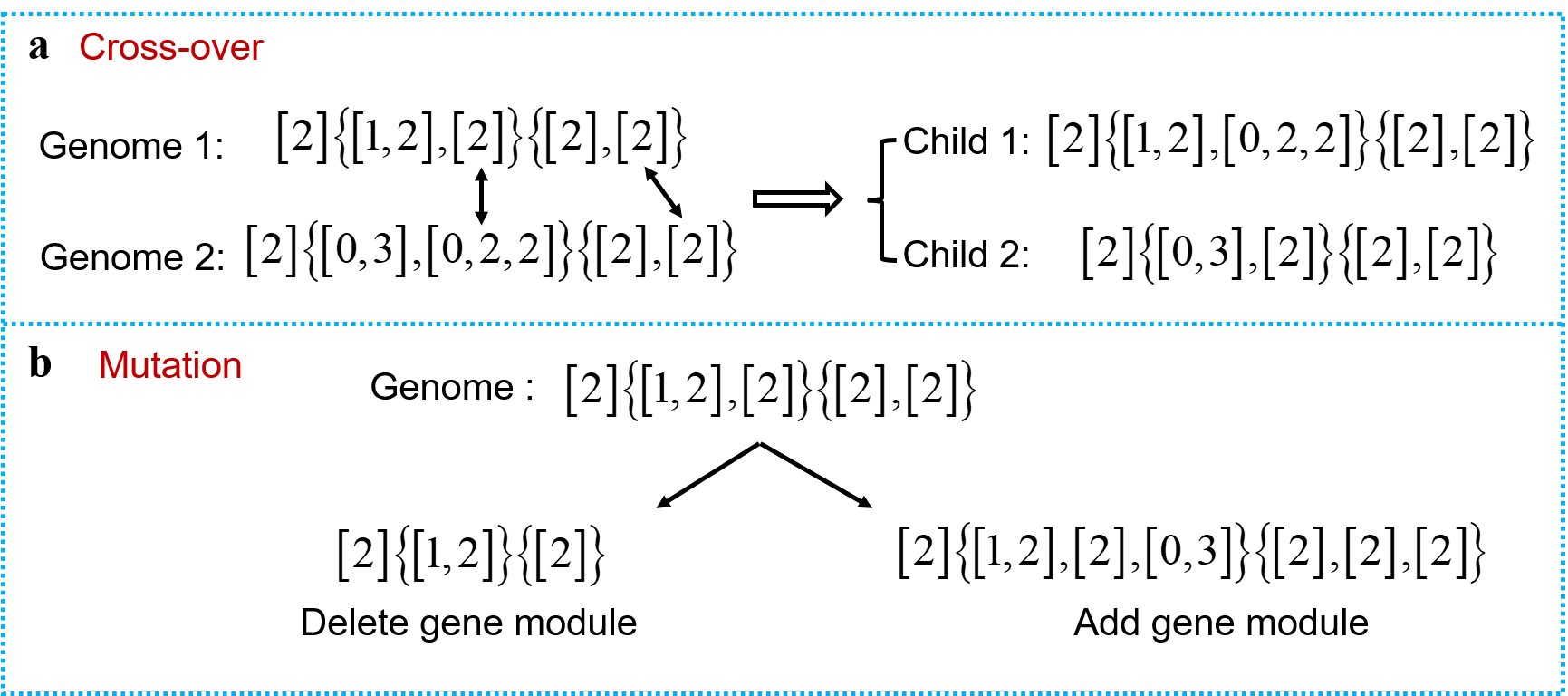}
\caption{An illustration of the process of cross-over and mutation. }
\label{fig2}
\end{figure}

Similar to the original framework, we use GA to evolve the library ${\Theta}(u)$ and its corresponding ${\Theta}(v)$ through multiple iterations. For each iteration, both libraries undergo the process of crossover, mutation, and selection sequentially. Compared to the original framework, these three operations are applied slightly different. For crossover, since we consider both functional and coefficient terms simultaneously, we swap functional and coefficient terms together, as illustrated in Figure \ref{fig2}a. The reason for this is that each functional and coefficient term correspond one-to-one, ensuring overall units conservation on the LHS of the equation. If we only swap functional or coefficient terms, the resulting equation may violate unit conservation. For mutation, we only use deletion and addition operations (see Figure \ref{fig2}b), eliminating order operations as they may cause unit inconsistency in the mutated genome. Also, deletion and addition of gene modules are performed simultaneously for both functional and coefficient terms since they are in a one-to-one correspondence. For selection, the fitness of each genome is calculated as follows:
\begin{equation}\label{eq13}
\mathcal{F} = \frac{1}{N}\sum\left(e q u^L-e q u_i^R \xi_i\right)^2 +\epsilon \cdot len(genome),
\end{equation}
where $N$ represents the total observed samples, $equ^L$ denotes the LHS functional terms of the potential wave equation, $equ_i^R$ stands for the product of the $i$th functional term and the corresponding coefficient term on the RHS, and the associated coefficients $\xi_i$ are determined through the application of singular value decomposition (SVD). The symbol $len(genome)$ denotes the length of the genome, and $\epsilon$ is a hyperparameter. A larger $\epsilon$ induces a simpler equation, while smaller values lead to a more complex form.

When the GA algorithm reaches the predetermined maximum number of iterations, we will inherit the approach of the original framework by using the optimal genome as a preliminary library. This approach has the advantage that, compared to a large library containing numerous candidate functional and coefficient terms where selecting the optimal combination from it is challenging, the preliminary library here contains only a limited number of combinations. Therefore, we can evaluate the redundancy and accuracy of each combination with a smaller computational cost using the more precise PIC criterion \citep{xu2023discovery}. Redundancy assessment relies on the moving horizon algorithm \citep{lejarza2022data}, while accuracy assessment is performed using PINN \citep{raissi2019physics} to calculate the physical loss. After obtaining the PIC score for each combination, the one with the minimum score will be selected as the discovered equation. Finally, to ensure that the numerical values on BHS of the equation are equal, we further utilize PINN to optimize the dimensionless coefficients in front of each functional and coefficient terms. With this, we complete the whole discovery process of a potential wave equation. Examples illustrating this discovery process will be presented in Section 4.1.

\section{Numerical Examples}
In the following, we take the discovery of the 2-D acoustic constant-density wave equation as an example to validate the effectiveness of our approach in discovering a physically interpretable wave equation.The classical acoustic wave equation has the following form:
\begin{equation}\label{eq14}
\setlength{\abovedisplayskip}{3pt}
\setlength{\belowdisplayskip}{3pt}
u_{t t}=v^2\left(u_{x x}+u_{z z}\right).
\end{equation}
Here, the body force is assumed as absent, and $v$ represents the velocity of the medium. The initial motion are activated by an isotropic Gaussian function as follows:
\begin{equation}\label{eq15}
\setlength{\abovedisplayskip}{3pt}
\setlength{\belowdisplayskip}{3pt}
u(i, k, 0)=\exp \left(-0.2 *\left[(i-x_0)^2+(k-z_0)^2\right]\right), \quad i=1, \cdots, N_x, k=1, \cdots, N_z,
\end{equation}
where $(x_0, z_0)$ denotes the center of Gaussian function, which is used to define the source location.

We start with discovering a wave equation in a homogeneous medium. We will illustrate the discovery process of our method, highlighting how the two libraries evolve and how we determine the equation's form. Subsequently, we evaluate the robustness of our approach to sparse and noisy observations. Moreover, we test our method's ability to discover wave equations in complex heterogeneous media. Finally, we consider a more realistic observation system to test the discovery capabilities of our method. All observed wavefields from these tests are simulated using finite-difference (FD) algorithm to solve Equation \ref{eq14}, with grid spacings of 10 $\texttt{m}$ in the $x$ and $z$ directions.

\subsection{The discovery process}
We consider wave propagation in a homogeneous medium with a velocity of 2 $\texttt{km}/\texttt{s}$ in an area of size $1\texttt{km} \times 1\texttt{km}$. The source is located at the center of the model. Subsequently, we record a total of 181 snapshots of the pressure wavefield from 0 to 0.36 $\texttt{s}$, with a time interval of 2 $\texttt{ms}$. We first randomly select a subset comprising 20\% of the complete volume of pressure wavefields to train the NN, approximating the selected wavefields. Here, we employ a simple fully-connected NN, using the sine function as the activation function. The network has 3 hidden layers, each with 256 neurons. The network is trained for 30000 epochs. The training of network is based on the Adam optimizer \citep{kingma2014adam}, with an initial learning rate of 1e-3, halved at the 5000th, 10000th, and 20000th epochs. This trained network will be used to represent the time and spatial derivatives in the GA, thereby translating the numerical encoding of potential equations into corresponding forms.

In the execution of the GA, the total number of genomes in the entire population library is set to 400, with a maximum number of generations of 100. As stated before, we simultaneously consider the cases of first- and second-order time derivatives on the LHS of the equation. The libraries for both cases evolve separately. Tables \ref{tab1} and \ref{tab2} display the optimal genomes at some generations, corresponding to the first- and second-order time derivatives on the LHS, respectively. The optimal genome is selected from the 400 genomes based on the fitness score, which is calculated using Equation \ref{eq13}. We can see that for the case of the first-order time derivative on the LHS, the GA reach a stable optimal genome by the 20th generations, where the coefficient $\xi_\text{5}$ corresponding to the terms $v(u_x^2+u_z^2)$ is relatively large, indicating its dominance in describing the wave propagation system. In contrast, for the case of the second-order time derivative on the LHS, the GA spends more generations to converge to the optimal genome. An interesting observation is that in all displayed generations, the coefficient corresponding to term $v^2(u_{xx}+u_{zz})$ is close to 1, significantly larger than the coefficients of other terms. This suggests that the GA identifies this term, which is present in the actual acoustic wave equation  used in modeling the observed data, while the other redundant terms have smaller coefficients and can be considered as fitting the observed wavefield values.

When the GA reaches the maximum specified number of generations, the optimal genomes for the first- and second-order time derivatives on the LHS of the equation constitute the preliminary libraries. We, then, iterate through all possible combinations of functional terms, which is finite. Subsequently, using the more accurate PIC criterion, we determine the optimal combinations, which represent the equation forms we discovered. Tables \ref{tab3} and \ref{tab4} list the equations with the five lowest PIC metrics, corresponding to equations with first- and second-order time derivatives on the LHS, respectively. We can find that equation ${u}_{tt}=0.922v^2(u_{xx} + u_{zz})$ has the lowest PIC metric, and thus, we determine it represents the discovered equation form.

However, we can see that there is still a certain discrepancy between the dimensionless coefficient value of 1 in front of the velocity terms of the real equation. This is because we only obtain an approximate value using SVD, which is not precise enough. Therefore, we embed the discovered equation (${u}_{tt}=\xi v^2(u_{xx} + u_{zz})$) into PINN, where the dimensionless coefficient $\xi$ is set as a learnable parameter. This parameter is initialized with the values obtained from SVD, that is, with the initial value of $\xi$ set to 0.922. Then, by iterating PINN 5000 epochs, we refine this dimensionless coefficient. Finally, we obtain the equation's ultimate form as ${u}_{tt}=0.988v^2(u_{xx} + u_{zz})$. It is evident that the discovered equation is very close to the accurate equation, with only minor differences in the dimensionless parameter. A part of the reason for this is that the observed wavefield is a numerical solution rather than an analytical solution. In other words, some minor dispersion might have preferred a lower velocity.

The above illustratation includes the entire process of discovering the wave equation from the observed wavefield. We demonstrated that our method can effectively discover a physically interpretable equation in a data-driven manner. Meanwhile, the cost of the discovery is relatively small, where the network training takes 780 s and the discovery process only spends 485 s.

\begin{table}
\centering
\caption{The evolution process of the optimal genome when the left-hand side of the equation is the first-order time derivative.}
\renewcommand\arraystretch{1.5}
\setlength{\tabcolsep}{16pt}
\begin{tabular}{ccc}
    \hline
     \multirow{3}{*} \text{Number of}  & \text{Optimal genome and } & \text {Fitness}  \\
     \text{generations}  & \text{Translation} & \\
    \hline
     $1$   & $\text {Optimal genome: } [1]\{[0, 2], [1, 1], [1, 1, 1]\}\{[1], [1], [1]\}$  &  $178.747$ \\
           & $\text {Translation: } \it{u}_{t}= \xi_\text{1}v(uu_{xx}+uu_{zz})+\xi_\text{2}v(u_x^2+u_z^2)+\xi_\text{3}v(u_x^3+u_z^3)$ &  \\
           & $\text {Coefficients: } \xi_\text{1} = -1.14\cdot10^{-3}, \quad \xi_\text{2} = 1.45\cdot10^{-2}, \quad \xi_\text{3} = -7.39\cdot10^{-6}$ &  \\
    \hline
    $20$     & $\text {Optimal Genome: } [1]\{[0, 0, 3], [1, 1, 1], [0, 2], [0, 1, 2], [1, 1]\}\{[1], [1], [1], [1], [1]\}$  &  $178.23$ \\
         & $\text {Translation: } \it{u}_{t}= \xi_\text{1}v(u^2u_{xxx}+u^2u_{zzz})+\xi_\text{2}v(u_x^3+u_z^3)$ &  \\
          & $ \quad \quad \quad  +\xi_\text{3}v(uu_{xx}+uu_{zz}) +  \xi_\text{4}v(uu_xu_{xx}+uu_zu_{zz})+\xi_\text{5}v(u_x^2+u_z^2)$ &  \\
         & $\text {Coefficients: } \xi_\text{1} = 9.91\cdot10^{-6}, \quad \xi_\text{2} = 4.41\cdot10^{-6}, $ & \\ 
         & $ \quad \quad \quad  \xi_\text{3} = -1.17\cdot10^{-3}, \quad \xi_\text{4} = -2.04\cdot10^{-5}, \quad \xi_\text{5} = 1.45\cdot10^{-2}$ \\
    \hline
    $40$     & $\text {Same with generation 20}$  &  $178.23$ \\
    \hline
    $60$     & $\text {Same with generation 20}$  &  $178.23$ \\
    \hline
    $80$     & $\text {Same with generation 20}$  &  $178.23$ \\
    \hline
    $100$     & $\text {Same with generation 20}$  &  $178.23$ \\
    \hline
\end{tabular}
\label{tab1}
\end{table}

\begin{table}
\centering
\caption{The evolution process of the optimal genome when the left-hand side of the equation is the second-order time derivative.}
\renewcommand\arraystretch{1.5}
\setlength{\tabcolsep}{11pt}
\begin{tabular}{ccc}
    \hline
     \multirow{3}{*} \text{Number of}  & \text{Optimal genome and } & \text {Fitness}  \\
     \text{generations}  & \text{Translation} & \\
    \hline
     $1$   & $\text {Optimal genome: } [2]\{[2], [1, 2], [0, 1, 3], [0, 2, 2]\}\{[2], [2], [2], [2]\}$  &  $86997.25$ \\
           & $\text {Translation: } \it{u}_{tt}= \xi_\text{1}v^2(u_{xx}+u_{zz})+\xi_\text{2}v^2(u_xu_{xx}+u_zu_{zz}) $ & \\
           & $+ \xi_\text{3}v^2(uu_xu_{xxx}+uu_zu_{zzz}) + \xi_\text{4}v^2(uu_{xx}^2+uu_{zz}^2)$ &  \\
           & $\text {Coefficients: } \xi_\text{1} = 9.96\cdot10^{-1}, \quad \xi_\text{2} = -3.13\cdot10^{-4}, $ & \\
           & $\quad \quad \quad \quad \quad \xi_\text{3} = -5.06\cdot10^{-6}, \quad \xi_\text{4} = -3.92\cdot10^{-6}$ &  \\
    \hline
    $20$     & $\text {Optimal Genome: } [2]\{[2], [0, 3], [0, 1, 3], [1, 2], [0, 2, 2]\}\{[2], [2], [2], [2], [2]\}$  &  $86963.8$ \\
           & $\text {Translation: } \it{u}_{tt}= \xi_\text{1}v^2(u_{xx}+u_{zz}) + \xi_\text{2}v^2(uu_{xxx}+uu_{zzz})$ & \\
           & $+ \xi_\text{3}v^2(uu_xu_{xxx}+uu_zu_{zzz}) + \xi_\text{4}v^2(u_xu_{xx}+u_zu_{zz}) + \xi_\text{5}v^2(uu_{xx}^2+uu_{zz}^2)$ &  \\
         & $\text {Coefficients: } \xi_\text{1} = 9.96\cdot10^{-1}, \quad \xi_\text{2} = 5.52\cdot10^{-5}, $ & \\ 
         & $ \quad \quad \quad  \xi_\text{3} = -5.17\cdot10^{-6}, \quad \xi_\text{4} = -3.52\cdot10^{-4}, \quad \xi_\text{5} = -3.88\cdot10^{-6}$ \\
    \hline
    $40$    & $\text {Optimal Genome: } [2]\{[2], [0, 1, 3], [1, 2], [1, 1, 2], [0, 2, 2]\}\{[2], [2], [2], [2], [2]\}$  &  $86881.59$ \\
           & $\text {Translation: } \it{u}_{tt}= \xi_\text{1}v^2(u_{xx}+u_{zz}) + \xi_\text{2}v^2(uu_xu_{xxx}+uu_zu_{zzz}) $ & \\
           & $+ \xi_\text{3}v^2(u_xu_{xx}+u_zu_{zz})  + \xi_\text{4}v^2(u_x^2u_{xx}+u_z^2u_{zz}) + \xi_\text{5}v^2(uu_{xx}^2+uu_{zz}^2)$ &  \\
         & $\text {Coefficients: } \xi_\text{1} = 9.95\cdot10^{-1}, \quad \xi_\text{2} = -4.69\cdot10^{-6}, $ & \\ 
         & $ \quad \quad \quad  \xi_\text{3} = -3.22\cdot10^{-4}, \quad \xi_\text{4} = 2.20\cdot10^{-6}, \quad \xi_\text{5} = -3.95\cdot10^{-6}$ \\
    \hline
    $60$     & $\text {Optimal Genome: } [2]\{[2], [0, 3], [1, 1, 2], [0, 1, 3], [0, 2, 2], [1, 2]\}\{[2], [2], [2], [2], [2], [2]\}$  &  $86847.47$ \\
           & $\text {Translation: } \it{u}_{tt}= \xi_\text{1}v^2(u_{xx}+u_{zz}) + \xi_\text{2}v^2(uu_{xxx}+uu_{zzz}) $ & \\
           & $+\xi_\text{3}v^2(u_x^2u_{xx}+u_z^2u_{zz}) + \xi_\text{4}v^2(uu_xu_{xxx}+uu_zu_{zzz}) $ & \\
           & $+ \xi_\text{5}v^2(uu_{xx}^2+uu_{zz}^2) + \xi_\text{6}v^2(u_xu_{xx}+u_zu_{zz}) $ &  \\
         & $\text {Coefficients: } \xi_\text{1} = 9.95\cdot10^{-1}, \quad \xi_\text{2} = 5.53\cdot10^{-5}, \quad \xi_\text{3} = -3.91\cdot10^{-6}, $ & \\ 
         & $ \quad \quad \quad \xi_\text{4} = -4.81\cdot10^{-6}, \quad \xi_\text{5} = -3.92\cdot10^{-6}, \quad \xi_\text{6} = -3.62\cdot10^{-4}$ \\
    \hline
    $80$     & $\text {Same with generation 60}$  &  $86847.47$ \\
    \hline
    $100$     & $\text {Same with generation 60}$  &  $86847.47$ \\
    \hline
\end{tabular}
\label{tab2}
\end{table}

\begin{table}
\centering
\caption{The potential wave equations and the corresponding PIC when the left-hand side of the equation is the first-order time derivative.}
\renewcommand\arraystretch{1.5}
\setlength{\tabcolsep}{20pt}
\begin{tabular}{cc} 
    \hline
    \text{Potential wave equation} & \text{PIC } \\
    \hline
    ${u}_{t}=0.0929v(u_x^2 + u_z^2)$ & $0.00622$ \\
    \hline
    ${u}_{t}=-0.0027v(u_x^3 + u_z^3) + 0.0964v(u_x^2 + u_z^2)$ & $0.00715$ \\
    \hline
    ${u}_{t}=-0.0260v(uu_{xx} + uu_{zz})$ & $0.00963$ \\
    \hline
    ${u}_{t}=0.0055v(u_x^3 + u_z^3) - 0.0052v(uu_{xx} + uu_{zz})$ & $0.01122$ \\
    \hline
    ${u}_{t}=0.0067v(u_x^3 + u_z^3)$ & $0.01297$ \\
    \hline
\end{tabular}
\label{tab3}
\end{table}

\begin{table}
\centering
\caption{The potential wave equations and the corresponding PIC when the left-hand side of the equation is the second-order time derivative.}
\renewcommand\arraystretch{1.5}
\setlength{\tabcolsep}{10pt}
\begin{tabular}{cc} 
    \hline
    \text{Potential wave equation} & \text{PIC } \\
    \hline
    ${u}_{tt}=0.922v^2(u_{xx} + u_{zz})$ & $0.0000939$ \\
    \hline
    ${u}_{tt}=0.0525v^2(u_xu_{xx} + u_zu_{zz})$ & $0.13459$ \\
    \hline
    ${u}_{tt}=0.06v^2(u_xu_{xx} + u_zu_{zz}) + 8.15\cdot10^{-6}v^2(uu_xu_{xxx} + uu_zu_{zzz}) $ & $0.15344$ \\
    \hline
    ${u}_{tt}=0.0589v^2(u_xu_{xx} + u_zu_{zz}) + 2.43\cdot10^{-5}v^2(uu_xu_{xxx} + uu_zu_{zzz}) - 5.75\cdot10^{-4}v^2(uu_{xxx} + uu_{zzz}) $ & $0.14821$ \\
    \hline
    ${u}_{tt}=0.0519v^2(u_xu_{xx} + u_zu_{zz}) + -0.0001v^2(uu_{xxx} + uu_{zzz})$ & $0.21739$ \\
    \hline
\end{tabular}
\label{tab4}
\end{table}

\subsection{Robust to noise and sparse observations}
\begin{table}
\centering
\caption{Test on discovery of a 2D acoustic wave equation with varying subsets of the total observations in a homogeneous medium.}
\renewcommand\arraystretch{1.5}
\setlength{\tabcolsep}{10pt}
\begin{tabular}{ccc}
    \hline
    \text {Volume} {of} {data} & \text { Discovered equation } & \text { Error } \\
    \hline
    $100 \%$ & ${u}_{tt}=0.985v^2 \left(u_{xx}+u_{zz}\right)$ & $1.5 \%$ \\
    $60 \%$ & ${u}_{tt}=0.985v^2 \left(u_{xx}+u_{zz}\right)$ & $1.5 \%$ \\
    $20 \%$ & ${u}_{tt}=0.988v^2 \left(u_{xx}+u_{zz}\right)$ & $1.2 \%$ \\
    $10 \%$ & ${u}_{tt}=0.983v^2 \left(u_{xx}+u_{zz}\right)$ & $1.7 \%$ \\
    $1 \%$ & ${u}_{tt}=1.048v^2 \left(u_{xx}+u_{zz}\right)$ & $4.8 \%$ \\
    $0.1 \%$ & ${u}_{tt}=0.967v^2\left(u_{xx}+u_{zz}\right)$ & $3.3 \%$ \\
    \hline
\end{tabular}
\label{tab5}
\end{table}

In real-world scenarios, the observed wavefield may be sparse. Also, due to environmental factors and sensor limitations, the collected wavefield may be contaminated by noise. Therefore, we validate the robustness of the proposed approach on noisy and sparse observations. Next, we will demonstrate our method's ability to discover wave equations from sparsely and noisy observed wavefields. 
We employ the same network architecture and training configuration, and also, use the wavefields simulated in the homogeneous medium from the previous section.

We first share the results of our method for discovering wave equations from sparsely observed wavefields. We randomly select subsets of pressure wavefields from all collected grid points as observed data, which serve as training data for the network. We choose subsets comprising 60\%, 20\%, 10\%, 1\%, and 0.1\% of all grid points, respectively, and compare the discovery results with those obtained using all grid points. Table \ref{tab5} presents the discovery results for varying subsets of pressure wavefields. We can observe that our method effectively discovers the complete form of the wave equation from observed data, while maintaining balanced units on BHS of the equation. To evaluate the deviation from the true equation (Equation \ref{eq14}), we use the relative error between the coefficient in front of the term $v^2$ and its real value of 1. We can see that our method maintains stable performance across different subsets of observations, with consistently small errors even for very limited observed data (e.g., 0.1\%). 

\begin{figure}
\centering
\includegraphics[width=1\textwidth]{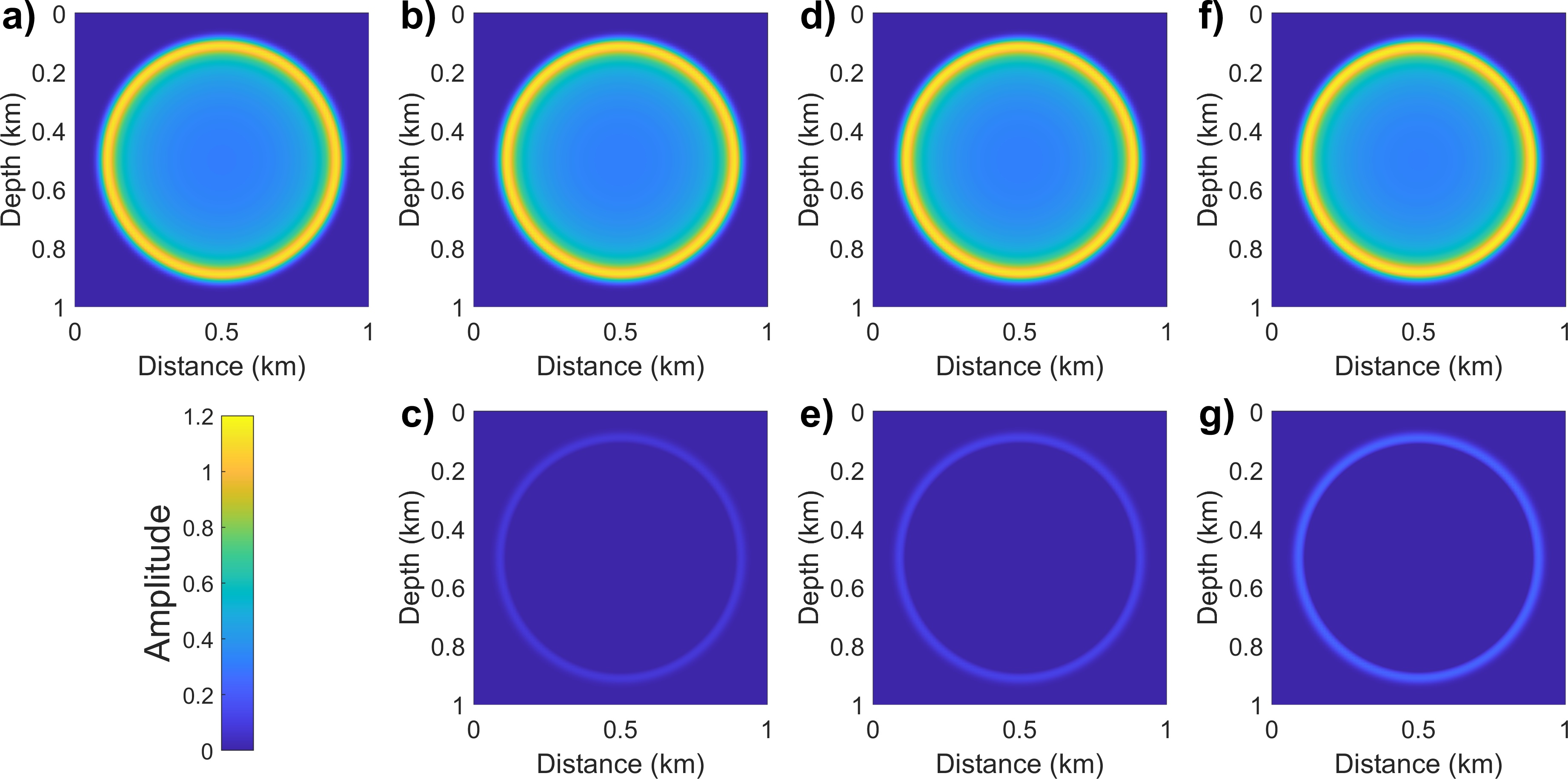}
\caption{Comparison of wavefield snapshots simulated by the accurate acoustic wave equation and the discovered equation with different data volumes. (a) Ground truth comes from accurate acoustic wave equation. (b), (d), and (f) are obtained by solving the discovered equations from $20 \%$, $10\%$, and $0.1\%$ volume data, respectively. (c), (e), and (g) are the corresponding differences with the ground truth.}
\label{fig3}
\end{figure}

We further employ FD algorithms to numerically solve the discovered equations derived from 20\%, 10\%, and 0.1\% volume data, respectively, and then compare these wavefields with those generated from simulating the real acoustic wave equation, as shown in Figure \ref{fig3}. We can observe that the wavefield snapshots simulated from the discovered equations are remarkably close to those derived from the accurate wave equation, exhibiting minimal residual differences. Even the wave equations identified from a mere 0.1\% volume data maintain exceedingly subtle discrepancies when compared to the true wavefields (see Figures \ref{fig3}f and g).

\begin{figure}
\centering
\includegraphics[width=1\textwidth]{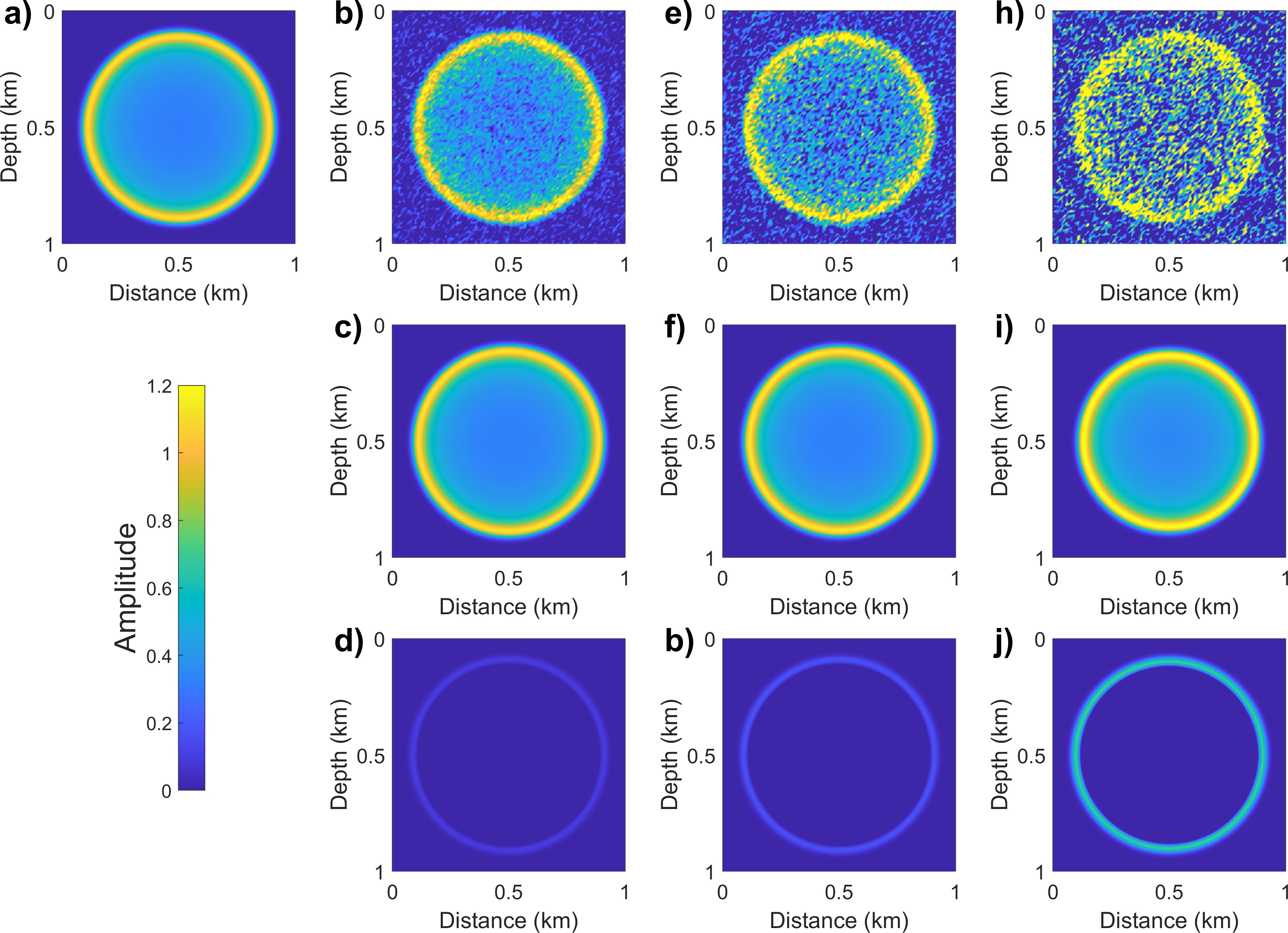}
\caption{Comparison of wavefield snapshots simulated by the accurate acoustic wave equation and the discovered equation with different noise levels. (a) Ground truth comes from accurate acoustic wave equation. (b), (e), and (h) are the noisy wavefield data with noise levels of $50\%$, $100\%$, and $200\%$, respectively, which are obtained by adding noise to the ground truth. (c), (f), and (i) are obtained by solving the discovered equations with noise levels of $50\%$, $100\%$, and $200\%$, respectively. (d), (g), and (j) are the corresponding differences with the ground truth.
}
\label{fig4}
\end{figure}

We, then, validate the robustness of our method to noisy observed wavefields. Here, we consider random noise. We generate noise using equation $\tilde{u}=u+\eta \cdot {std}\left(u \right) \cdot N\left(0,1 \right)$, where $N\left(0,1 \right)$ denotes the standard normal distribution with mean 0 and standard deviation of 1, and $\eta$ is the noise level. We randomly select a subset comprising 20\% of the simulated data from all grid points and inject noise into them to create noisy observations. Table \ref{tab6} shows the discovery results of our method for observations with different noise levels, including 25\%, 50\%, 100\%, 200\%, 300\%, and 400\%. We observe that for moderate noise levels, such as 25\% to 100\%, the discovered equations exhibit very low errors. As the noise level increases, the error of our method gradually increases. However, the forms of the functional terms on BHS of the discovered equation, as well as the coefficient term, remain consistent with the true wave equation. At a noise level of 400\%, our method discovers an incorrect equation, although the units on BHS of the equation remain balanced. This outcome is anticipated because such strong noise can obscure the wavefield, leading to fitting noise rather than the true wavefield. As a result, it significantly impacts the network's ability to interpolate sparse wavefields and the accuracy of computing spatial and time derivatives.

We also solve the equations discovered from the data with noise levels of 50\%, 100\%, and 200\%, respectively. Figure \ref{fig4} shows a comparison between the generated wavefield snapshots and their corresponding ground truth. We can see that our method can provide a accurate equation within low noise. As noise increases, there are more errors, but the results are still commendable. With very high noise, it's hard to observe clear waveform, but the discovered equations still manage to produce wavefield close to the true solutions.

\begin{table}
\centering
\caption{Test on discovery of a 2D acoustic wave equation from data with different noise level in a homogeneous medium.}
\renewcommand\arraystretch{1.4}
\setlength{\tabcolsep}{10pt}
\begin{tabular}{ccc}
    \hline
    \text {Noise} {level} & \text { Discovered equation } & \text { Error } \\
    \hline
    $25 \%$ & ${u}_{tt}=0.987v^2 \left(u_{xx}+u_{zz}\right)$ & $1.3 \%$ \\
    $50 \%$ & ${u}_{tt}=0.986v^2 \left(u_{xx}+u_{zz}\right)$ & $1.4 \%$ \\
    $100 \%$ & ${u}_{tt}=0.976v^2 \left(u_{xx}+u_{zz}\right)$ & $2.4 \%$ \\
    $200 \%$ & ${u}_{tt}=0.894v^2 \left(u_{xx}+u_{zz}\right)$ & $10.6 \%$ \\
    $300 \%$ & ${u}_{tt}=0.658v^2 \left(u_{xx}+u_{zz}\right)$ & $34.2 \%$ \\
    $400 \%$ & ${u}_{t}=-0.002v\left(uu_{xx}+uu_{zz}\right)$ &  \\
    \hline
\end{tabular}
\label{tab6}
\end{table}

\subsection{Discovery in inhomogeneous media}
\begin{figure}
\centering
\includegraphics[width=0.5\textwidth]{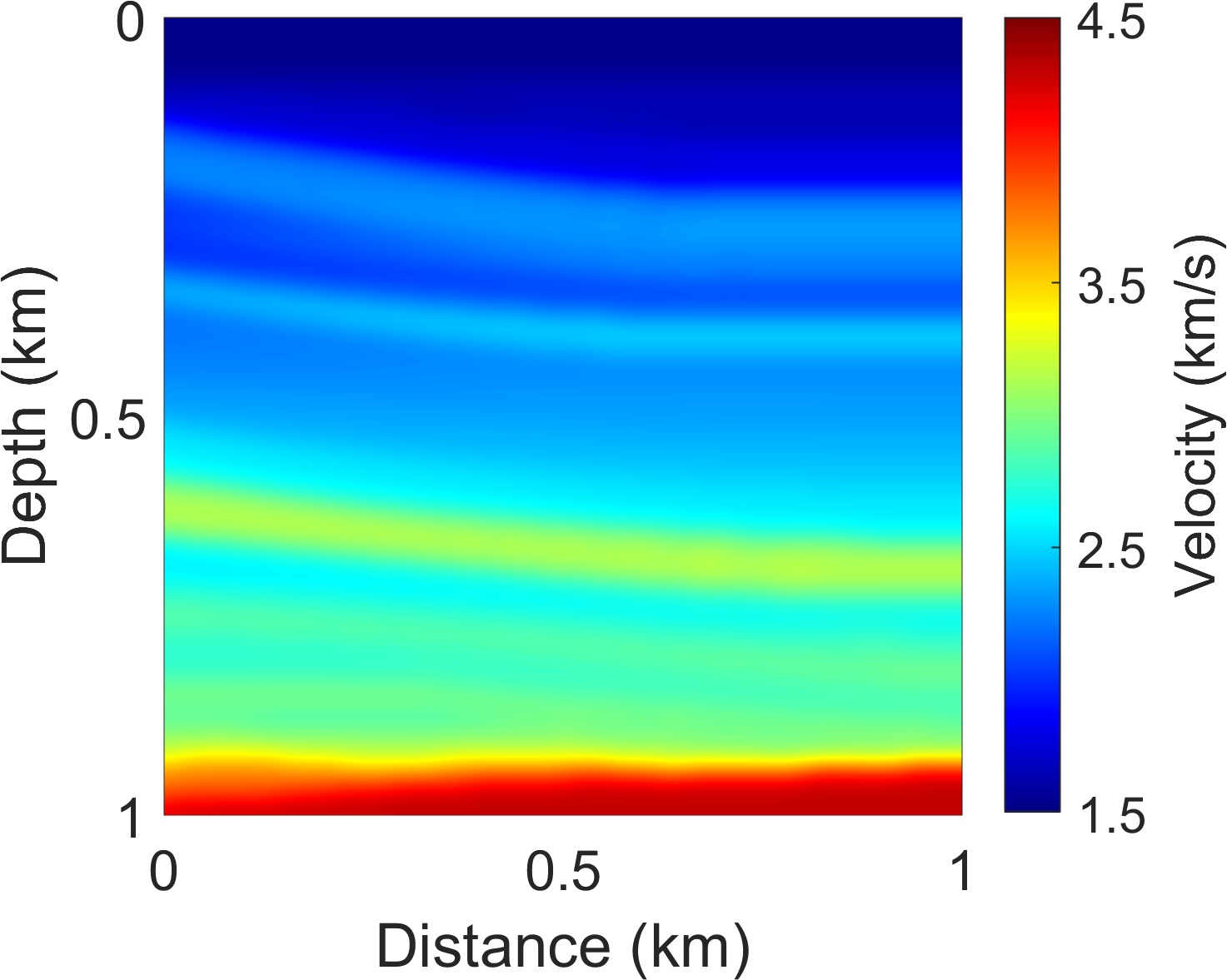}
\caption{The velocity model.}
\label{fig5}
\end{figure}

We further present the discovery results of our method in a more complex inhomogeneous medium. Figure \ref{fig5} shows the tested velocity model, sized  1 $\texttt{km} \times$1 $\texttt{km}$, which is extracted from the Marmousi model. Similarly, we place the source at the model's center and then utilize FD to solve the acoustic wave equation to obtain 201 pressure wavefield snapshots from 0 to 4 $\texttt{s}$, with a time interval of 2 $\texttt{ms}$. We randomly select 20\% of the complete volume of the wavefields as training observed data to consider sparse observation scenarios. Meanwhile, we inject random noise with levels of 50\%, 100\%, 200\%, and 300\% to the selected sparse observations, thereby assessing our method's capability to discover wave equations for noisy wavefields in inhomogeneous media. The network share the same architecture and training configuration as in the Section 4.1.

Table \ref{tab7} presents the discovery results for both clean (noise level = 0\%) and noisy sparse observations. This table illustrates that our method is capable of effectively identifying the forms of wave equations under common noise levels, ranging from 0\% to 200\%. Also, it reveals a trend where increasing noise in the wavefield leads to a gradual deviation of the dimensionless coefficients on the RHS of the discovered equations from the expected value of 1, which we also observe in homogeneous media. In comparison to our discovery in homogeneous media, where our method successfully identified wave equations in noisy wavefields up to 400\% noise level, here, in this inhomogeneous medium, an incorrect wave equation is identified at a 300\% noise level. This discrepancy is attributed to the increased complexity of wavefields, which include additional wave phenomena such as reflections and transmissions.

\begin{figure}
\centering
\includegraphics[width=1\textwidth]{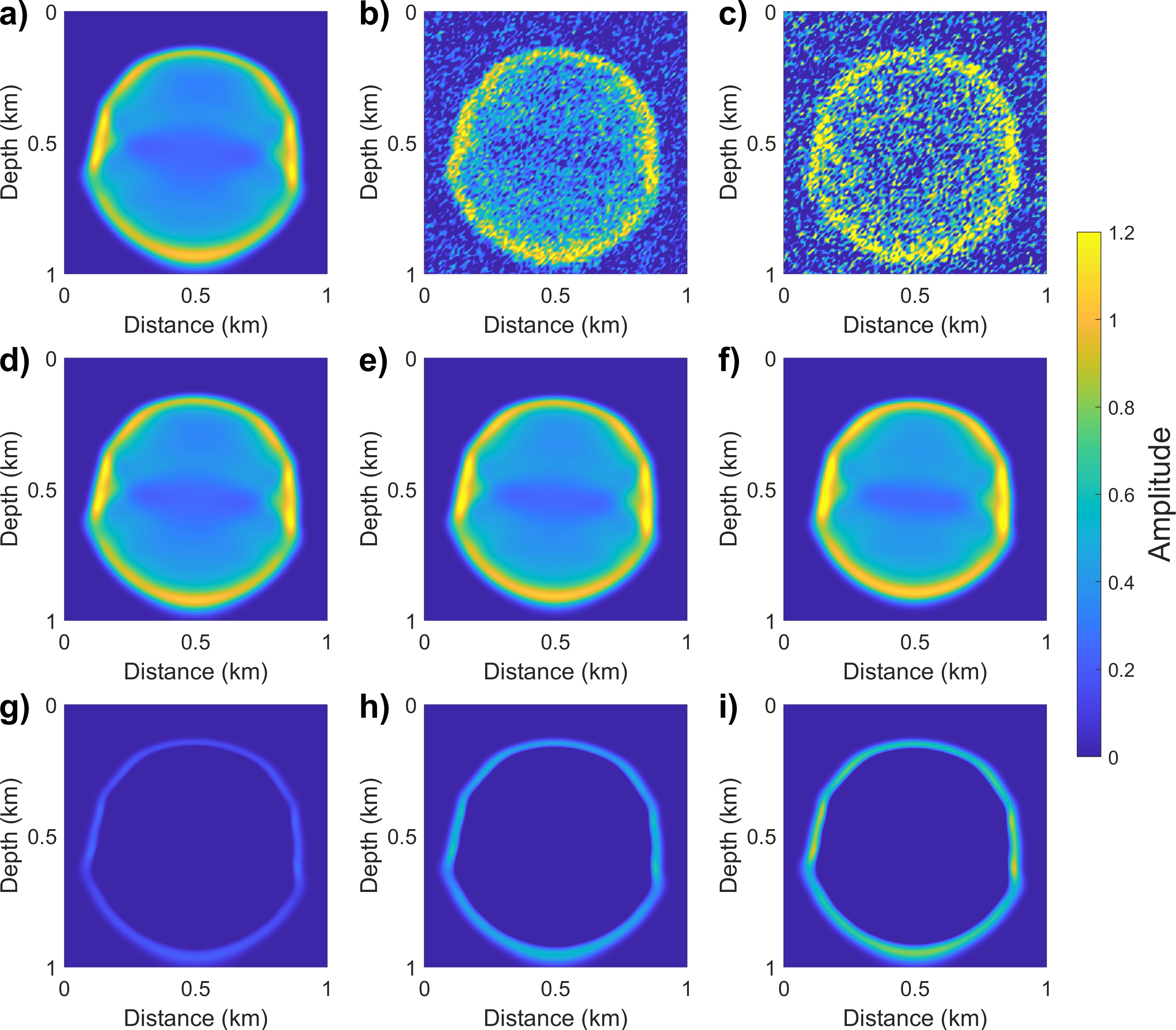}
\caption{Comparison of wavefield snapshots simulated by the accurate acoustic wave equation and the discovered equation in an inhomogeneous medium. (a) Ground truth comes from accurate acoustic wave equation. (b) and (c) are the noisy wavefield data with noise levels of $100\%$ and $200\%$, respectively, which are obtained by adding noise to the ground truth. (d), (e), and (f) are obtained by solving the discovered equations from $20\%$ volume wavefield and those with noise levels of $100\%$ and $200\%$, respectively. (g), (h), and (i) are the corresponding differences with the ground truth.
}
\label{fig6}
\end{figure}

We numerically solve the discovered equations from clean sparse data and noisy data at noise levels of 100\% and 200\%. Figure \ref{fig6} displays the wavefield snapshots from these discovered equations. Panel a shows the accurate wavefield snapshot, while panels b and c depict the accurate wavefield with added random noise at levels of 100\% and 200\%, respectively, to illustrate the contamination in the noisy observations used for discovery. Panels d, e, and f correspond to the wavefield snapshots from the discovered equations using clean sparse data and noisy data at 100\% and 200\% noise levels, respectively. Panels g, h, and i show the differences between these resulting wavefields and ground truth. We can observe that the wavefields induced by the discovered equations hold a significant degree of consistency with the ground truth from the accurate acoustic wave equation. Naturally, as the noise increases, such as in panel \ref{fig6}c, the discrepancy between the discovered and ground truth increases. This is expected since the observed wavefield is severely contaminated at higher noise levels.

\begin{figure}
\centering
\includegraphics[width=1\textwidth]{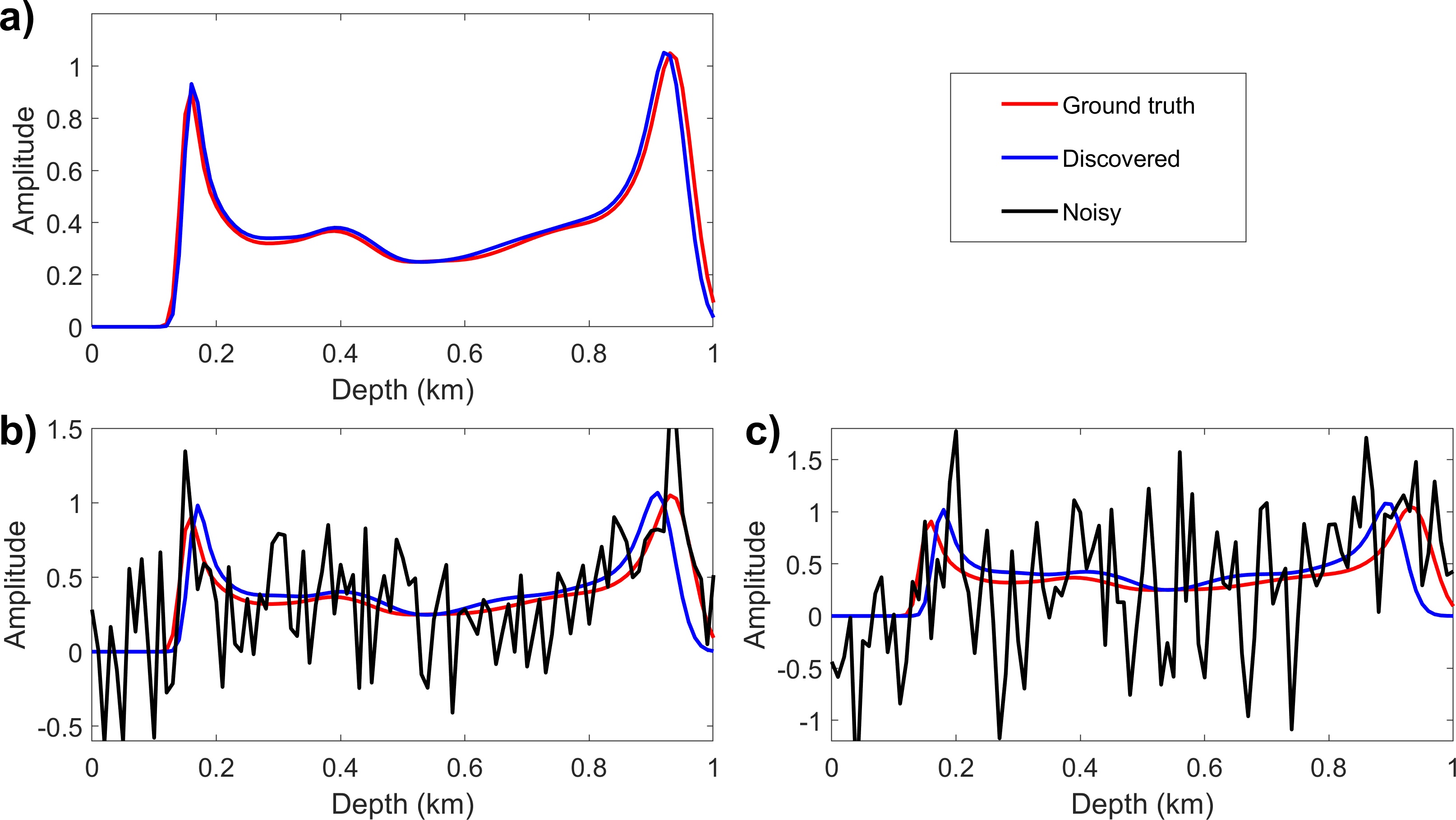}
\caption{Comparison of single-trace waveforms simulated by the accurate acoustic wave equation (red line) and the discovered equation (blue line) in an inhomogeneous medium, where the trace is located at a horizontal distances $x=0.5$ \texttt{km}. Black line represents the noisy observations used for discovery. (a) Discovered equation from clean observations. (b) and (c) Discovered equations from the noisy observations with noise levels of $100\%$ and $200\%$, respectively.}
\label{fig7}
\end{figure}

To more clearly show the differences in amplitude and phase between waveforms simulated by the discovered equations and the accurate equations, we compare single-trace waveforms at horizontal location $x=0.5$ \texttt{km} in Figure \ref{fig7}. Panel a corresponds to the equation discovered from clean sparse wavefields, while panels b and c correspond to equations discovered from noisy wavefields at noise levels of 100\% and 200\%, respectively. In panel a, we can see that, in the absence of noise pollution, the waveforms simulated by our discovered equation closely match the ground truth in both phase and amplitude. In panels b and c, it is evident that the phase and amplitude of the waveforms from the noisy wavefields (shown by the black line) are significantly distorted. This illustrates the fundamental challenge of discovering equations from noisy wavefields. Nonetheless, the equations discovered by our method still manage to provide wavefield solutions that are relatively close to the ground truth.

\begin{table}
\centering
\caption{Test on discovery of a 2D acoustic wave equation from data with different noise level in an inhomogeneous medium.}
\renewcommand\arraystretch{1.4}
\setlength{\tabcolsep}{10pt}
\begin{tabular}{ccc}
    \hline
    \text {Noise} {level} & \text { Discovered equation } & \text { Error } \\
    \hline
    $0 \%$ & ${u}_{tt}=0.968v^2 \left(u_{xx}+u_{zz}\right)$ & $3.2 \%$ \\
    $50 \%$ & ${u}_{tt}=0.953v^2 \left(u_{xx}+u_{zz}\right)$ & $4.7 \%$ \\
    $100 \%$ & ${u}_{tt}=0.912v^2 \left(u_{xx}+u_{zz}\right)$ & $8.8 \%$ \\
    $200 \%$ & ${u}_{tt}=0.860v^2 \left(u_{xx}+u_{zz}\right)$ & $14 \%$ \\
    $300 \%$ & ${u}_{t}=0.019v \left(u_{x}^2+u_{z}^2\right)$ & \\
    \hline
\end{tabular}
\label{tab7}
\end{table}

\subsection{Discovery in realistic observations}
Although we have demonstrated that we can effectively discover wave equations from sparse and noisy data, the observed wavefields used were randomly selected from all grid points. Actually, it is not feasible to measure wavefields internally within a model. In most cases, a more realistic observation system would only collect wavefields at the surface of a medium (or along its boundaries), as is done in seismic surveys. For this reason, we will consider a more realistic observation setup moving forward.

We use the homogeneous medium from Section 3.1 as an example. We place the receivers only on its top surface while we also place the source in the middle of the model's top surface. All receivers record from 0 to 0.57 $\texttt{s}$ with a time interval of 1 $\texttt{ms}$. Under such restricted observations, an NN with a large number of neurons could lead to severe overfitting, resulting in significant errors in wavefield interpolation and the estimation of partial derivatives. Therefore, we reduce the number of neurons per layer from 256 to 56.

Table \ref{tab8} displays the discovery results for wavefields at different noise levels under this observation system. We can see that, even under such limited observations, our method can identify the form of the acoustic wave equation from noise levels ranging from 0 to 200\%. Naturally, compared to the randomly selected observed wavefields used in the previous sections, the dimensionless coefficients of the equations discovered here show greater deviation from the true value of 1. This deviation is acceptable considering that we are utilizing very limited observed wavefields to guide our discovery. When the noise level reaches 300\%, our method discovers an incorrect equation. Interestingly, under this observation setup, a clean wavefield did not induce the discovery of the most accurate equation. Instead, a smaller error is found in the equation discovered at a noise level of 200\%. This finding contrasts with earlier conclusions. A possible reason might be that the distribution of the randomly generated noise at this level and the wavefield at the boundaries are more aligned, thereby minimally impacting the optimization of dimensionless coefficients using PINN.

\begin{table}
\centering
\caption{Test on discovery of a 2D acoustic wave equation in realistic observations with different noise level.}
\renewcommand\arraystretch{1.4}
\setlength{\tabcolsep}{10pt}
\begin{tabular}{ccc}
    \hline
    \text {Noise} {level} & \text { Discovered equation } & \text { Error } \\
    \hline
    $0 $ & ${u}_{tt}=1.177v^2 \left(u_{xx}+u_{zz}\right)$ & $17.7 \%$ \\
    $50 \%$ & ${u}_{tt}=1.162v^2 \left(u_{xx}+u_{zz}\right)$ & $16.2 \%$ \\
    $100 \%$ & ${u}_{tt}=1.165v^2 \left(u_{xx}+u_{zz}\right)$ & $16.5 \%$ \\
    $200 \%$ & ${u}_{tt}=1.078v^2 \left(u_{xx}+u_{zz}\right)$ & $7.8 \%$ \\
    $300 \%$ & ${u}_{tt}=-0.999v^2 \left(uu_xu_{xxx}+uu_zu_{zzz}\right)$ &  \\
    \hline
\end{tabular}
\label{tab8}
\end{table}

\section{Discussion}
\subsection{Benefits of physical unit constraints}
If we merely employ regression methods to fit numerical values of observed wavefields, the resulting discovered equations may only be applicable to wave propagation systems within a single medium. When the physical properties of the medium change, wave behaviors can undergo significant alterations. As a result, the discovered equations would be unable to describe the dynamic behavior of new media. In the context of physics theory, any mathematical equation describing wave phenomena must adhere to a hard constraint: the physical units on both-hand sides (BHS) of the equation must be balanced. This balance provides an elegant formal aspect to physical equations. Thus, any equation, to be physically interpretable and useful for describing wave propagation, must satisfy this requirement. This principle ensures that the equations derived from observations are not just mathematically accurate representations of specific data sets but also generalize across different conditions by maintaining physical validity. This physical interpretability is crucial for the equations to be reliably applied in different contexts, particularly when extrapolating to conditions not directly observed during the data collection phase.

Leveraging this distinctive feature of physical equations, we proposed the use of physical unit constraints to guide the discovery of wave equations. This constraint is utilized to direct the generation of potential candidate functional and coefficient terms within the library. If the combination of randomly generated functional and coefficient terms does not conserve the units on BHS of the equation, it is immediately discarded and not included in the library. This strategy offers two significant benefits.

On the one hand, incorporating physical unit constraints significantly reduces the search space. For example, in our framework, we consider both first- and second-order time derivatives on the left-hand side (LHS) of the equation. When the LHS involves a first-order time derivative, according to our unit encoding and counting principle (see Section 3.3), both units $\texttt{m}$ and $\texttt{s}$ are counted as 1. To maintain units balance, both units on the right-hand side (RHS) must also hold 1. On the RHS, only the coefficients of the terms include unit $\texttt{s}$, and also, the coefficients of the terms, where the unit $\texttt{s}$ is equal to 1, is confined to $v$. This implies that when the LHS of the equation features a first-order time derivative, the coefficient on the RHS must exclusively be $v$. This is why all coefficient terms shown in Table \ref{tab1} are $v$. Furthermore, since the unit $\texttt{m}$ for the coefficient term $v$ is defined as 1, the unit $\texttt{m}$ for any functional term paired with $v$ must be 0 to keep the unit $\texttt{m}$ balanced on BHS of the equations. This requirement restricts the optimization space to the range where the unit $\texttt{m}$ of the functional terms on the RHS must be 0. Consequently, we effectively reduce this optimization task from potentially infinite candidate functional terms to a limited target space.

On the other hand, the equations we discover will conserve units, thus possessing the potential to be physically interpretable. To substantiate this argument, we conduct a comparative test by removing the unit constraints during the discovery process. Table \ref{tab9} presents the discovery results for varying subsets of pressure wavefields without unit constraints. While the form of the functional terms on BHS of the equation can still be discovered, we end up with an incorrect form of the coefficient term, leading to imbalanced units on BHS of the equation. As a result, it can not serve as a physical model for studying wave phenomena.

\begin{table}
\centering
\caption{Test on discovery of a 2D acoustic wave equation with varying subsets of the total observations without using unit constraints.}
\renewcommand\arraystretch{1.4}
\setlength{\tabcolsep}{20pt}
\begin{tabular}{cc}
    \hline
    \text {Volume} {of} {data} & \text { Discovered equation }  \\
    \hline
    $100 \%$ & ${u}_{tt}=0.246v^4 \left(u_{xx}+u_{zz}\right)$  \\
    $60 \%$ & ${u}_{tt}=0.246v^4 \left(u_{xx}+u_{zz}\right)$ \\
    $20 \%$ & ${u}_{tt}=0.247v^4 \left(u_{xx}+u_{zz}\right)$  \\
    $10 \%$ & ${u}_{tt}=0.524v^3 \left(u_{xx}+u_{zz}\right)$  \\
    $1 \%$ & ${u}_{tt}=0.523v^3 \left(u_{xx}+u_{zz}\right)$  \\
    $0.1 \%$ & ${u}_{tt}=0.242v^4\left(u_{xx}+u_{zz}\right)$  \\
    \hline
\end{tabular}
\label{tab9}
\end{table}

\subsection{Future work}
In our previous \citep{cheng2023d, cheng2024robust}, as well as this work, we proposed a data-driven approach to directly discover wave equations from observed wavefields. Taking the acoustic wave equation as an example, we explored the discovery of its corresponding mathematical form from simulated pressure wavefields. Preliminary numerical experiments demonstrated the reliability of this theory. How, then, might future work further explore its potential applications in seismological research?

To the best of our knowledge, we believe there are two research directions worth further exploration:
\begin{itemize}
   \item  Using our framework to discover new equations. We can see that our current work still manages to discover an existing equation. However, a huge potential of this work is to discover new equations. As we stated earlier, traditional methods always rely on existing physical laws to derive wave equations. However, we demonstrated the feasibility of a data-driven discovery method through numerical experiments. Therefore, we can combine this discovery approach with laboratory rock physics measurements. By utilizing measurements of wavefields, dispersion, or attenuation from rock physics, we can directly discover new wave equations or rock physics equations that might not yet be known or fully understood. This can be combined with the use of lab scaled physical models in which we properties of the model is known, and we use measurements to discovery the corresponding wave equation. This can include physical models with anisotropy and attenuation \citep{alkhalifah2003acoustic, hao2021nearly, wang2022propagating}.

   \item Using a data-driven approach to simplify equations. Due to the complexity of wave physics, some wave equations have very intricate explicit forms to describe wave propagation. These complex forms can lead to challenging numerical implementations and computational burdens. Alternatively, we can simulate observed wavefields using their complex forms, and then, utilize our method to attempt to discover their corresponding simplified forms, thereby significantly reducing the difficulty of numerical implementations and computational cost.
   
\end{itemize}
\section{Conclusions}
We presented a significant enhancement in the data-driven discovery of wave equations through the innovative incorporation of physical unit constraints and a reoptimized algorithm. Building upon the original algorithm, we considered encoding the coefficient terms to enable the resulting equations to provide specific forms for both the functional and coefficient terms. Also, we employed the hard constraint that physical models must have balanced units, and thus, to ensure that all generated potential equations are physically interpretable. This strategy eliminates the generation of non-physical equations and significantly reduces the search space for potential equations, thereby guiding the discovery process towards accurate wave equations. 

We tested the new algorithm for discovering the 2-D acoustic wave equation. The results demonstrated that our method effectively discovers the accurate and complete form of wave equation from noisy and limited observed wavefields in both homogeneous and inhomogeneous media, even in the presence of noise up to certain levels. Meanwhile, we demonstrated that our method can effectively discover wave equations under a realistic observation system, such as seismic surveys that record wavefields only at the surface. All tests underscore the critical importance of physical unit balance, ensuring that the discovered equations are not only mathematically robust but also physically meaningful. This significant advancement provides a robust foundation for future explorations into more nuanced and realistic scenarios.
\bmhead{Acknowledgments}
This publication is based on work supported by the King Abdullah University of Science and Technology (KAUST). The authors thank the DeepWave sponsors for supporting this research. This work utilized the resources of the Supercomputing Laboratory at King Abdullah University of Science and Technology (KAUST) in Thuwal, Saudi Arabia.
\bmhead{Code and Data Availability}
The accompanying codes that support the findings will be shared upon acceptance of this paper.

\backmatter

\section*{Declarations}
\textbf{Confict of interest} We declare: “No confict of interest exists in the submission of this manuscript, and manuscript is approved by all authors for publication. We would like to declare on behalf of my co-authors that the work described was original research that has not been published previously, and not under consideration for publication elsewhere, in whole or in part. All the authors listed have approved the manuscript that is enclosed.” \\

\bibliography{references}

\end{document}